 \documentclass[final,5p,times,twocolumn]{elsarticle}

\usepackage{amssymb}
\usepackage{amsmath}
\newcommand{\Lagr}{\mathcal{L}}
\newcommand{\no}{\nonumber}
\newcommand{\ba}{\begin{eqnarray}}
\newcommand{\ea}{\end{eqnarray}}
\def\bra#1{\left\langle #1\right|}
\def\ket#1{\left| #1\right\rangle}

\begin{document}

\begin{frontmatter}



\title{The $\Xi(1620)$ and $\Xi(1690)$ molecular states from $S=-2$ meson-baryon interaction up to next-to-leading order.}


\author[inst1]{A. Feijoo}

\affiliation[inst1]{organization={Departamento de Física Teórica and IFIC, Centro Mixto Universidad de Valencia-CSIC},
            addressline={Institutos de Investigación de Paterna, Aptdo.~22085}, 
            city={46071 Valencia},
            country={Spain}}

\author[inst2]{V. Valcarce Cadenas}
\author[inst2,inst3]{V.K. Magas}

\affiliation[inst2]{organization={Departament de Física Quàntica i Astrofísica, Universitat de Barcelona},
            addressline={Martí Franquès 1},
            city={08028 Barcelona},
            country={Spain}}

\affiliation[inst3]{organization={Institut de Ciències del Cosmos, Universitat de Barcelona},
            addressline={Martí Franquès 1}, 
            city={08028 Barcelona},
            country={Spain}}

\begin{abstract}
We have studied the meson-baryon interaction in the neutral $S=-2$ sector using an extended Unitarized Chiral Perturbation Theory, which takes into account not only the leading Weinberg-Tomozawa term (as all the previous studies in $S=-2$ sector), but also the Born terms and next-to-leading order contribution. Based on the SU(3) symmetry of the chiral Lagrangian we took most of the model parameters from the BCN model \cite{Feijoo:2018den}, where these were fitted to a large amount of experimental data in the neutral $S=-1$ sector. We have shown that our approach is able to generate dynamically both $\Xi(1620)$ and $\Xi(1690)$ states in very reasonable agreement with the data, and can naturally explain the puzzle with the decay branching ratios of $\Xi(1690)$.  Our results clearly 
 illustrate the reliability of chiral models implementing unitarization in coupled channels and the importance of considering Born and NLO contributions for precise calculations.  
\end{abstract}




\end{frontmatter}


\section{Introduction}
\label{sec:introduction}
The recent analysis of the $\Xi^-_b \to J/\psi \Lambda K^-$ decay by the LHCb Collaboration showed the presence of two narrow excited $\Xi^-$ states in the $K^- \Lambda$ invariant mass distribution \cite{LHCb:2020jpq}. On a one to one assignation, the peak located at lower energy was identified as the $\Xi(1690)^-$ while the second one as the $\Xi(1820)^-$. 
Their masses and widths were measured with improved precision in comparison with \cite{ParticleDataGroup:2022pth}, where the typical uncertainties are about $5$ MeV. 
The corresponding data treatment and fitting procedure established the $\Xi(1690)^-$ mass and width as:
\begin{equation}
M=1692.0\pm1.3^{+1.2}_{-0.4}~\text {MeV},\, \Gamma=25.9\pm9.5^{+14.0}_{-13.5}~\text {MeV}.
\label{eq:Xi_1690}
\end{equation}

The LHCb measurement was preceded by the first observation of the $\Xi(1620)^0$ baryon decaying into $\Xi^-\pi^+$ via the $\Xi^+_c \to \pi^+\pi^+\Xi^-$ process, which was collected with the Belle detector at the KEKB asymmetric-energy $e^+e^-$ collider \cite{Belle:2018lws}. The subsequent data analysis determined its mass and width as:
\begin{equation}
M=1610.4\pm6.0^{+5.9}_{-3.5}~\text {MeV},\, \Gamma=59.9\pm4.8^{+2.8}_{-3.0}~\text {MeV}.
\label{eq:Xi_1620}
\end{equation}
In addition, the $\Xi^-\pi^+$ invariant mass distribution studied in \cite{Belle:2018lws} shed some more light on the structure of hyperon resonances with strangeness $S=-2$ by providing a clear signal of the $\Xi(1530)^0$ resonance and a structure that can be attributed to the $\Xi(1690)^0$ state.

With all previous measurements, and bearing in mind that the information reported by LHCb in \cite{LHCb:2020jpq} has not been yet updated in the PDG compilation \cite{ParticleDataGroup:2022pth}, the $\Xi(1620)$ state is rated with a single star status and its spin and parity ($J^P$) need to be confirmed. Actually, it has been common practice to assume $J^P={1/2}^-$ as the corresponding spin-parity given the analogy with the $\Lambda(1405)$ as its counterpart in $S=-1$. Apparently, the status of the $\Xi(1690)^0$ seems to be clearer since it is rated with three-star and its spin-parity is established as $J^P={1/2}^-$ from an experimental evidence in the $\Lambda^+_c \to K^+\pi^+\Xi^-$ decay \cite{BaBar:2008myc}.

On the theoretical side, there is still a long-term controversy about the nature of the $\Xi(1620)$ and $\Xi(1690)$ states. There are certain indications pointing to the fact that these states may have a nontrivial internal structure rather than a plain $qqq$ configuration. On the one hand, as already mentioned, the unavoidable analogy between $\Xi(1620)$ and $\Lambda(1405)$ leads one to interpret the $\Xi(1620)$ as a molecular state arising from the Unitarized Chiral Perturbation (UChPT) scheme \cite{ParticleDataGroup:2022pth,Jido:2003cb}. On the other hand, according to the study of \cite{BaBar:2008myc}, the spin-parity of the $\Xi(1690)$ should be $J^P={1/2}^-$, a fact that qualifies such a state to decay into the $\pi\Xi$, $\bar K\Lambda$ or $\bar K\Sigma$ channels in s-wave. By inspecting the $\Xi(1690)$ branching ratios in \cite{ParticleDataGroup:2022pth}, one can appreciate that the $\Gamma_{\pi\Xi}/\Gamma_{\bar K\Sigma}$ one is less than $0.09$ in spite of the fact that $\pi\Xi$ has a much larger phase space than $\bar K\Sigma$. This issue was already addressed in \cite{Sekihara:2015qqa}, where the $\Xi(1690)$ was interpreted as a $\bar K\Sigma$ quasibound state dynamically generated from the Chiral Unitary approach in coupled channels \cite{Kaiser:1995eg,Oset:1997it}. The explanation for the smallness of the above branching ratio was found to be due to the nearly vanishing $\pi\Xi$ couplings. 

 Being chronologically accurate, the $S=-2$ meson-baryon interaction within the SU$(3)$ UChPT approach in coupled channels was first employed  in \cite{Ramos:2002xh}, the authors of which found one resonance that was assigned to the $\Xi(1620)$ state ruling out any possibility of being the $\Xi(1690)$ state. This work was followed by a similar study \cite{Garcia-Recio:2003ejq}, and some of the previous authors revisited the meson-baryon interaction in $S=0,-1,-2,-3$ sectors yet using a SU$(6)$ extension of the Chiral Lagrangian within a coupled channel unitary approach in \cite{Gamermann:2011mq}. Among the large number of resonances found in these two last studies, the $\Xi^*$ poles with $J^P={1/2}^-$ describe qualitatively the properties of the $\Xi(1620)$, $\Xi(1690)$ and $\Xi(1950)$ states. In \cite{Sekihara:2015qqa}, the $\Xi(1690)$ was dynamically generated in good agreement with experimental mass yet with a tiny width. Regarding the $\Xi(1620)$ state, only one virtual state was found at an energy more than $50$ MeV below the experimental location. Recently, a new study \cite{Nishibuchi:2022zfo} based on the same approach as the one in \cite{Ramos:2002xh} pinned the $\Xi(1620)$ down to the experimental value at the expense of reducing unnaturally one of the parameters present in the unitarization method. 
 
The forthcoming experiments studying hadronic decays of charm baryons, governed by $c \to s$ transitions, will provide more information about the $S=-2$ baryon spectroscopy. The ongoing measurements of $\gamma p \to K^+ (\Lambda^*/\Sigma^*) \to K^+K^{+/0}\Xi^{*-/0}(\to K^+K^{+/0} \bar K \Lambda ) $ by GlueX Collaboration \cite{Pauli:2022ehd} can also play a crucial role on this issue. Furthermore, the $K^- \Lambda$ Correlation Function can be experimentally accessed with Femptoscopy Technique by ALICE at LHC \cite{ALICE:2020wvi} as it was done for $\bar KN$ \cite{ALICE:2019gcn,ALICE:2022yyh}.

 In view of that, the current theoretical models should be revisited and improved. In this sense, in the present work, we make a step forward and, for the first time in this sector, we take into account higher order contributions in the Lagrangian from which the interaction kernel is derived. In all previous theoretical works \cite{Sekihara:2015qqa, Kaiser:1995eg, Oset:1997it, Ramos:2002xh, Garcia-Recio:2003ejq, Gamermann:2011mq, Nishibuchi:2022zfo}, only a contact Weinberg-Tomozawa (WT)-like term was used as  interaction. Further perturbative corrections have been systematically ignored since they are assumed to play a very moderate role, specially, in s-wave. And, so far, it has been seen that these models provide a plausible explanation on the nature of these $\Xi^*$ states in terms of meson-baryon molecules. However, we wonder how strong this assumption is and whether these higher order terms can help to describe accurately the mass and width of such states simultaneously, what was not achieved in the above cited papers. 

\begin{figure}[ht]
\begin{center}
\centering
\includegraphics[width=3.4 in]{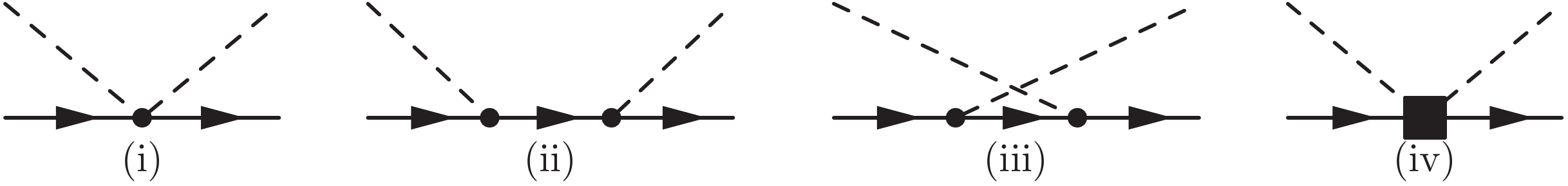}
\vspace{-0.4cm}
\caption{\label{fig:epsart} Feynman diagrams for meson-baryon interaction: WT term (i), direct and crossed Born terms (ii) and (iii), and NLO terms (iv). Dashed (solid) lines represent the pseudoscalar octet mesons (octet baryons).}
\end{center}
\end{figure}
 
 In the $S=-1$ sector by looking at $\bar KN$ interaction we can found evidences of the non-negligible role of the s- and u-channels, known as direct and cross Born terms,  and the tree level next-to-leading order (NLO) contribution, schematically shown in Fig. \ref{fig:epsart}. For instance, in \cite{Oller:2000fj}, the authors pointed out that the Born contributions reach $20\%$ of the dominant WT contribution just $65$ MeV above the $\bar KN$ threshold. As the energy increases moderately, the combined effect between the Born and the NLO terms play a crucial role in the reproduction of the total cross section from $\bar KN \to \eta\Lambda,\eta\Sigma,K\Xi$ processes \cite{Feijoo:2015yja,Ramos:2016odk,Feijoo:2018den}. At this point, it should be recalled that the value of the $\eta$-channel thresholds are located a little bit more than $200$ MeV above $\bar KN$ threshold. Similarly,  the difference between $\bar K^0 \Lambda$ and the $\eta \Xi^0$ threshold is around $250$ MeV. Thus, in the $S=-2$ sector a similar impact of these terms can be expected in the energy regime delimited by the higher thresholds. These new pieces in the interaction kernel will enable processes that are not connected with the WT term and, consequently, the additional interplay among the channels may affect the widths and the locations of the dynamically generated states.
 
With all this in mind, in the present study we have incorporated to the meson-baryon interaction in the neutral $S=-2$ sector, in addition to WT term, the s-,u-channel Born diagrams and the tree level NLO contribution,  by adapting the BCN model (WT+Born+NLO model in \cite{Feijoo:2018den}). And we show that such an extended model is able to generate dynamically both $\Xi(1620)$ and $\Xi(1690)$ states in a fair agreement with experimental data.  

\section{Formalism}
\label{sec:formalism}
The derivation of the meson-baryon interaction from effective chiral lagrangians has been addressed many times in literature \cite{Kaiser:1995eg,Oset:1997it,Lutz:2001yb}. Precisely for this reason, in this section we only highlight the main steps and underline the particularities of our model. As a starting point, the $SU(3)$ chiral effective Lagrangian up to NLO is taken into consideration,  
%
$
\Lagr_{\phi B}^{eff}=\Lagr_{\phi B}^{(1)}+\Lagr_{\phi B}^{(2)}  \ ,
$
%
with $\Lagr_{\phi B}^{(1)}$ and $\Lagr_{\phi B}^{(2)}$ being the most general form of the leading order and NLO contributions to meson-baryon interaction Lagrangian, respectively, defined as  
\ba 
\Lagr_{\phi B}^{(1)} & = & i \langle \bar{B} \gamma_{\mu} [D^{\mu},B] \rangle
                            - M_0 \langle \bar{B}B \rangle  
                           - \frac{1}{2} D \langle \bar{B} \gamma_{\mu} 
                             \gamma_5 \{u^{\mu},B\} \rangle \no \\
                  & &      - \frac{1}{2} F \langle \bar{B} \gamma_{\mu} 
                               \gamma_5 [u^{\mu},B] \rangle \ ,
\label{LagrphiB1} 
\ea 
\begin{eqnarray}
    \Lagr_{\phi B}^{(2)}& = & b_D \langle \bar{B} \{\chi_+,B\} \rangle
                             + b_F \langle \bar{B} [\chi_+,B] \rangle
                             + b_0 \langle \bar{B} B \rangle \langle \chi_+ \rangle \no \\ 
                     &  & + d_1 \langle \bar{B} \{u_{\mu},[u^{\mu},B]\} \rangle 
                            + d_2 \langle \bar{B} [u_{\mu},[u^{\mu},B]] \rangle    \no \\
                    &  &  + d_3 \langle \bar{B} u_{\mu} \rangle \langle u^{\mu} B \rangle
                            + d_4 \langle \bar{B} B \rangle \langle u^{\mu} u_{\mu} \rangle \ .
\label{LagrphiB2}
\end{eqnarray}

The $3\times3$ unitary matrix $B$ contains the fundamental baryon octet $(N,\Lambda,\Sigma,\Xi)$, in both equations. The incorporation of the pseudoscalar meson octet ($\pi,K,\eta$) requires a more complicated prescription, $u_\mu = i u^\dagger \partial_\mu U u^\dagger$, to preserve the chiral symmetry. The pseudoscalar fields are collected in a $3\times3$ unitary $\phi$  matrix, which enters via $U(\phi) = u^2(\phi) = \exp{\left( \sqrt{2} {\rm i} \phi/f \right)} $, where $f$ is the meson decay constant. In Eqs.~(\ref{LagrphiB1}) and (\ref{LagrphiB2}), the symbol $\langle \dots \rangle$ stands for trace in flavor space. 

The Low Energy Constants (LECs) $D$ and $F$ present in Eq.~(\ref{LagrphiB1}) are the so called $SU(3)$ axial vector constants, which are subjected to the constraint $g_A=D+F=1.26$. In the same expression, $M_0$ is the common baryon octet mass in the chiral limit. The  covariant derivative,  $[D_\mu, B] = \partial_\mu B + [ \Gamma_\mu, B]$ with $\Gamma_\mu =  [ u^\dagger,  \partial_\mu u] /2$ being the chiral connection, is present in Eq.~(\ref{LagrphiB1}). Furthermore, in Eq.~(\ref{LagrphiB2}), we find $\chi_+ = 2 B_0 (u^\dagger \mathcal{M} u^\dagger + u \mathcal{M} u)$, which breaks chiral symmetry explicitly via the quark mass matrix  $\mathcal{M} = {\rm diag}(m_u, m_d, m_s)$ and $B_0 = - \bra{0} \bar{q} q \ket{0} / f^2$; the latter is related to the order parameter of spontaneously broken chiral symmetry. The corresponding LECs at NLO, namely $b_D$, $b_F$, $b_0$ and $d_i$ $(i=1,\dots,4)$, need to be determined from experiment since they are not fixed by the symmetries of the underlying theory. Given the lack of scattering data availability in the $S=-2$ sector, we effectively assume the $SU(3)$ symmetry in the present work, and thus we use the LECs obtained in the $S=-1$  sector for the WT+Born+NLO model (see Table~(II) in \cite{Feijoo:2018den}).

\begin{table*}[h]
\begin{center}
\resizebox{17.4cm}{!} {
\begin{tabular}{|c|cccccc|}
\hline
& & & & & & \\[-1.5mm]
$D_{ij}$ & {\bf $\pi^+\Xi^-$} & {\bf $\pi^0\Xi^0$} & {\bf $\bar{K}^0\Lambda$} & {\bf $K^-\Sigma^+$} & {\bf $\bar{K}^0\Sigma^0$} & {\bf $\eta\Xi^0$} \\ [2mm] \hline
& & & & & & \\[-1.5mm]
{\bf $\pi^+\Xi^-$} & $2(2b_0+b_D-b_F)m^{2}_{\pi}$ & $0$ & ${-(b_D-3b_F)\mu_{1}^{2}}/{\sqrt{6}}$ & $0$  & ${(b_D+b_F)\mu_{1}^{2}}/{\sqrt{2}}$  & ${2\sqrt{2}(b_D-b_F)m^{2}_{\pi}}/{\sqrt{3}}$  \\ [2mm]
{\bf $\pi^0\Xi^0$}   &  &$4(b_0+b_D-b_F)m^{2}_{\pi}$ &${(b_D-3b_F)\mu_{1}^{2}}/{2\sqrt{3}}$  & ${(b_D+b_F)\mu_{1}^{2}}/{\sqrt{2}}$ &  ${(b_D+b_F)\mu_{1}^{2}}/{2}$  & ${-2(b_D-b_F)m^{2}_{\pi}}/{\sqrt{3}}$  \\ [2mm]
{\bf $\bar{K}^0\Lambda$} &    &    &  ${2(6b_0+5b_D)m^{2}_{K}}/{3}$  &  ${2\sqrt{2}b_{D}m^{2}_{K}}/{\sqrt{3}}$ & ${-2b_{D}m^{2}_{K}}/{\sqrt{3}}$ &  ${(b_D-3b_F)\mu^{2}_{2}}/{6}$  \\ [2mm]
{\bf $K^-\Sigma^+$}&    &    &      & $2(2b_0+b_D+b_F)m^{2}_{K}$ & $-2\sqrt{2}b_{F}m^{2}_{K}$  & ${-(b_D+b_F)\mu^{2}_{2}}/{\sqrt{6}}$ \\ [0mm]
{\bf $\bar{K}^0\Sigma^0$}  &     &    &      &       & $2(2b_0+b_D)m^{2}_{K}$  & ${(b_D+b_F)\mu^{2}_{2}}/{2\sqrt{3}}$  \\ [1mm]
{\bf $\eta\Xi^0$} &    &     &     &     &   & ${2(2b_0\mu^{2}_{3}+b_D\mu^{2}_{4}+b_F\mu^{2}_{5})}/{3}$ \\ [2mm] \hline
\multicolumn{7}{c}{} \\[1mm]
\hline
& & & & & & \\[-1.5mm]
$L_{ij}$ & {\bf $\pi^+\Xi^-$} & {\bf $\pi^0\Xi^0$} & {\bf $\bar{K}^0\Lambda$} & {\bf $K^-\Sigma^+$} & {\bf $\bar{K}^0\Sigma^0$} & {\bf $\eta\Xi^0$} \\ [2mm] \hline
& & & & & & \\[-1.5mm]
{\bf $\pi^+\Xi^-$} & $-d_1+d_2+2d_4$ & $0$ & ${\sqrt{3}(d_1-d_2)}/{\sqrt{2}}$ & $-2d_2+d_3$  & ${(d_1+3d_2)}/{\sqrt{2}}$  & ${-\sqrt{2}(d_1-3d_2)}/{\sqrt{3}}$  \\ [2mm]
{\bf $\pi^0\Xi^0$}   &  &$-d_1+d_2+2d_4$ &${-(d_1-3d_2)}/{2\sqrt{3}}$  & ${(d_1+d_2)}/{\sqrt{2}}$ &  ${-(d_1+d_2)}/{2}$  & $0$  \\ [2mm]
{\bf $\bar{K}^0\Lambda$} &    &    &  ${(d_1+3d_2+2d_4)}/{2}$  &  $\sqrt{6}d_2$ & $-\sqrt{3}d_2$ &  $0$  \\ [2mm]
{\bf $K^-\Sigma^+$}&    &    &      & $0$ & $-\sqrt{2}d_1$  & ${-(d_1+3d_2)}/{\sqrt{6}}$ \\ [2mm]
{\bf $\bar{K}^0\Sigma^0$}  &     &    &      &       & $d_2+2d_4$  & ${(d_1+3d_2)}/{2\sqrt{3}}$  \\ [2mm]
{\bf $\eta\Xi^0$} &    &     &     &     &   & $d_1+3d_2+2d_4$ \\ [2mm] \hline
\end{tabular}
}
\caption{$D_{ij}$ and $L_{ij}$ coefficients in the NLO potential of the pseudoescalar meson and the baryon octet with strangeness $S = -2$ and charge $Q = 0$. The coefficients are symmetric, $D_{ji} = D_{ij}$ and $L_{ji} = L_{ij}$. With the definitions: $\mu_{1}^{2} = m_{K}^{2} + m_{\pi}^{2}$, $\mu_{2}^{2} = 5m_{K}^{2} - 3m_{\pi}^{2}$, $\mu_{3}^{2} = 4m_{K}^{2} - m_{\pi}^{2}$, $\mu_{4}^{2} = 8m_{K}^{2} - 3m_{\pi}^{2}$, $\mu_{5}^{2} = 8m_{K}^{2} - 5m_{\pi}^{2}$.}
\label{D_L}
\end{center}
\end{table*}

In  Fig.~\ref{fig:epsart}, the different contributions to the meson-baryon interaction kernel are diagrammatically represented. More precisely, the contact diagram (i) corresponds to the WT contribution; this comes from the term with the covariant derivative in Eq.~(\ref{LagrphiB1}). Next, the vertices of diagrams (ii) and (iii), which stand for the direct and crossed Born contributions, are obtained from the second and third terms of Eq.~(\ref{LagrphiB1}). And finally, the contribution of the NLO contact diagram, i.e. the fourth diagram of Fig.~\ref{fig:epsart}, is directly extracted from Eq. (\ref{LagrphiB2}).

Thus, the total interaction kernel up to NLO can be expressed as the sum of all those terms: 
%
 $
 V_{ij}=V^{\scriptscriptstyle WT}_{ij}+V^{direct}_{ij}+V^{crossed}_{ij}+V^{\scriptscriptstyle NLO}_{ij}\,,
 $
%
where the elements of the $V_{ij}= \langle i | V | j \rangle $ interaction matrix couple the meson-baryon channels which, in the present case, amount to six: $\pi^0\Xi^0$, $\pi^+\Xi^-$, $\bar{K}^0 \Lambda$, $K^-\Sigma^+$, $\bar{K}^0 \Sigma^0$ and $\eta\Xi^0$. The explicit expressions for each contribution to the interaction kernel (for $S=-1$ sector) can be found in Eqs.~(7)-(10) of \cite{Feijoo:2021zau} that, once projected onto s-wave, they recover the structure showed by Eqs.~(6)-(8) and (10) in \cite{Ramos:2016odk}. The expressions for $S=-2$ are just the same, but obviously the corresponding coefficients should be recalculated for this sector. Actually, the $C_{ij}$ coefficients of the WT contact potential can be found in Table~1 of Ref.~\cite{Sekihara:2015qqa}. The baryon-meson-baryon coefficients present in the pseudoscalar-coupling vertexes in both Born diagrams can be obtained taking into account the relations (A.5) in Appendix~A of Ref.~\cite{Borasoy:2005ie}. Finally, one can find the NLO Clebsch-Gordan type coefficients, $D_{ij}$ and $L_{ij}$ in Eq.~(10) of \cite{Ramos:2016odk}, displayed in Table~\ref{D_L}.

Once the $V_{ij}$ potential is calculated, the scattering matrix is obtained by solving the Bethe-Salpetter (BS) equation in coupled channels by factorizing the interaction kernel and the scattering amplitude out of the integral equation, leaving a simple system of algebraic equations to be solved which, in matrix form, reads as (see \cite{Feijoo:2018den} and the references therein for a more detailed explanation)   
%
$
T_{ij} ={(1-V_{il}G_l)}^{-1}V_{lj}\,,
$
%
where $T_{ij}$ represents the scattering amplitude for a given starting i-channel and an outgoing j-channel, and $G_l$ is the loop function standing for a diagonal matrix with elements: 
\begin{equation} \label{Loop_integral}
G_l={\rm i}\int \frac{d^4q_l}{{(2\pi)}^4}\frac{2M_l}{{(P-q_l)}^2-M_l^2+{\rm i}\epsilon}\frac{1}{q_l^2-m_l^2+{\rm i}\epsilon} ,
\end{equation} 
where $M_l$ and $m_l$ are the baryon and meson masses of the $l$-channel. The dimensional regularization is applied on this function because of its logarithmic divergence to finally get:
\ba
& G_l = &\frac{2M_l}{(4\pi)^2} \Bigg \lbrace a_l(\mu)+\ln\frac{M_l^2}{\mu^2}+\frac{m_l^2-M_l^2+s}{2s}\ln\frac{m_l^2}{M_l^2} + \no \\ 
 &     &\frac{q_{\rm cm}}{\sqrt{s}}\ln\left[\frac{(s+2\sqrt{s}q_{\rm cm})^2-(M_l^2-m_l^2)^2}{(s-2\sqrt{s}q_{\rm cm})^2-(M_l^2-m_l^2)^2}\right]\Bigg \rbrace .  
 \label{dim_reg}    
\ea 
The loop function $G_l$ comes in terms of the subtraction constants (SC) $a_l$ that replace the divergence for a given dimensional regularization scale $\mu$, which is taken to be $630$~MeV in the present work. These constants are unknown, however, one can establish a natural size for them following \cite{Oller:2000fj}, and it comes out to be around $-2.0$; in our study we will allow the substracting constants to vary within the interval $[-3.5,-1]$.  
In addition, isospin symmetry arguments are frequently used to reduce the number of independent SC, in particular, we consider $4$ such constants here ($a_{\pi\Xi}$, $a_{\bar{K}\Lambda}$,$a_{\bar{K}\Sigma}$ and $a_{\eta\Xi}$). 

The dynamically generated resonance states show up as pole singularities of the scattering amplitude at $\sqrt{s}=z_p=M_R-{\rm i}\Gamma_R/2$, whose real and imaginary parts correspond to its mass ($M_R$) and the half width ($\Gamma_R/2$). The complex coupling strengths ($g_i$, $g_j$) of the resonance to the corresponding meson-baryon channels can be evaluated assuming a Breit-Wigner structure for the scattering amplitude in the proximity of the found pole on the real axis,
%
$
T_{ij}(\sqrt{s})\sim {g_i g_j} / {(\sqrt{s}-z_p)} \,.
$
%

\section{Results}
\label{sec:results}

Our starting point is a model derived from a chiral Lagrangian up to NLO in s-wave (WT+Born+NLO model of \cite{Feijoo:2018den}), which involves a number of parameters, LECs plus SC, that amount to a maximum of $16$ in the $S=-1$ sector. In the sector of interest, $S=-2 (Q=0)$, due to a smaller number of available channels a number of parameters is reduced to $14$, namelly: the meson decay constant $f$, the axial vector couplings $D$ and $F$ , the NLO coefficients $b_0$, $b_D$, $b_F$, $d_1$, $d_2$, $d_3$, $d_4$; and four SC $a_{\pi\Xi}$, $a_{\bar{K}\Lambda}$, $a_{\bar{K}\Sigma}$ and $a_{\eta\Xi}$. In the present work, as first step, all the LECs are assumed to be $SU(3)$ symmetric, meaning that the values of $f$, $D$, $F$, $b_0$, $b_D$, $b_F$, $d_1$, $d_2$, $d_3$ and $d_4$ are fixed to the ones obtained in \cite{Feijoo:2018den}. Thus, the strategy followed consists of varying the SC within a reasonable natural size range, $[-3.5, -1]$, in order to describe the properties of the $\Xi(1620)$ and $\Xi(1690)$ states in the best possible way. This procedure is called Model I in the present study.

\begin{table}[h]
  \caption{Values of the parameters for the different models described in the text. The subtraction constants are taken at a regularization scale $\mu=630$~MeV.} 
  \label{tab:outputs_fits}
\centering
\begin{tabular}{lcc}
\hline \\[-2.5mm]
                                 & \textbf{Model I}             &  \textbf{Model II}      \\                              
\hline \\[-2.5mm]

$a_{\pi\Xi}$                   & $-2.7981$                         & $ -2.7228$        \\
$a_{\bar{K}\Lambda}$           & $ -1.0071$                        & $ -1.0000 $       \\
$a_{\bar{K}\Sigma}$            & $-3.0938$                         &  $ -2.9381$       \\
$a_{\eta\Xi}$                  & $ -3.2665$                        &  $ -3.3984 $      \\
$f/f_{\pi}$                    & $1.197$  \, (fixed \cite{Feijoo:2018den})   &  $ 1.204$               \\
\hline
\end{tabular}
\end{table}

It is worth mentioning that, prior to proceed with the full interaction kernel, we take the leading order contact term (WT) as the only contribution in the interaction kernel and tried to reproduce the $\Xi^*$ of interest by means of tuning the SC. As found in the literature mentioned above, one is able to dynamically generate two poles, but the features of those could not reproduce the experimental masses and widths of $\Xi(1620)$ and $\Xi(1690)$ at the same time. 
First of all,  for WT-based models even if two resonances are generated there exists a mutual incompatibility in pinning both masses down simultaneously.
Furthermore, the width of the $\Xi(1690)$ remains invariably small around $1$~MeV. The $\Xi(1620)$ resonance can be reproduced very well, but only allowing  some of the SC to take unnatural-size values. 

By contrary, the situation rather improves when all contributions are included in the kernel. This is clearly seen by inspecting the results for Model I in Table~\ref{tab:spectroscopy} using the corresponding parametrization from Table~\ref{tab:outputs_fits}. The most eye-catching output is the closeness of both masses to the experimental values simultaneously. In particular, the energy location of the pole associated to $\Xi(1620)$ is within the experimental error band (Eq.~(\ref{eq:Xi_1620})) while the pole related to $\Xi(1690)$ is about $7$~MeV below the lower edge of the corresponding experimental error band (Eq.~(\ref{eq:Xi_1690})). Furthermore, the value for the theoretical width of $\Xi(1690)$ is within the error band (Eq.~(\ref{eq:Xi_1690})). This is in contrast to what happens to the theoretical $\Xi(1620)$ width that reaches about a factor $2.5$ bigger than the nominal value (Eq.~(\ref{eq:Xi_1620})) (this issue is addressed below). 

As a next step, in Model II  we decided to consider the global scaling factor $f$ as a free parameter of the model, yet keeping the constraint that this effective decay constant can only take values within the corresponding error bands of the BCN model (Table~II of Ref.~\cite{Feijoo:2018den}). Actually, the value of the constant $f$ in our model does not correspond exactly to the known experimental value of $f_\pi$, because it effectively takes into account the role of kaons and  $\eta$'s  in the system. This is teh case not only for  Ref.~\cite{Feijoo:2018den}, but also for some other studies were UChPT models are fitted to the data - effectively $f$ is a bit larger than $f_\pi$.
However, the role of each pseudoscalar meson in the $S=-2$ sector differs from the one in the $S=-1$ sector, studied in \cite{Feijoo:2018den}, therefore it is reasonable to readjust the $f$ value for the current study. 

\begin{table}[t]
  \caption{Comparison of the pole positions between the models: Model I and Model II (in MeV) with their couplings  $g_i$ and the corresponding modulus found in $J^P=\frac{1}{2}^-$, $(I,S)=(\frac{1}{2},-2)$.}
  \smallskip
  \label{tab:spectroscopy}
\centering
\begin{tabular}{c|cc|cc}
\hline
\hline 
 {\bf  Model I} &    \multicolumn{2}{c|}{$\Xi(1620)$}    &  \multicolumn{2}{c}{$\Xi(1690)$} \\
 \hline
  
$M\;\rm[MeV]$           & \multicolumn{2}{c|}{$1599.95$}      & \multicolumn{2}{c}{$1683.04$}     \\
$\Gamma\;\rm[MeV]$      & \multicolumn{2}{c|}{$158.88$}       & \multicolumn{2}{c}{$11.51$}   \\
\hline
                         &   $ g_i$          &   $|g_i|$      &  $g_i$      & $|g_i|$        \\
$\pi \Xi$                &  $2.09+i1.00$     &  $2.32$         & $-0.30-i0.12$   &  $0.33$   \\
$\bar{K}\Lambda$         &  $-2.11-i0.09$    &  $2.11$         & $-0.49+i0.05$  &  $0.50$   \\
$\bar{K}\Sigma$          & $-0.90+i0.34$     &  $0.97$         &   $1.57-i0.24$  &  $1.59$   \\
$\eta \Xi$               &  $-0.23+i0.13$    &  $0.26$         &  $0.74-i0.11$    & $0.74$   \\
\hline
\hline
{\bf  Model II} &    \multicolumn{2}{c|}{$\Xi(1620)$}    &  \multicolumn{2}{c}{$\Xi(1690)$} \\
 \hline
$M\;\rm[MeV]$     & \multicolumn{2}{c|}{$1608.51$}   & \multicolumn{2}{c}{$1686.17$}     \\
$\Gamma\;\rm[MeV]$   & \multicolumn{2}{c|}{$170.00$}   & \multicolumn{2}{c}{$29.72$}   \\
\hline
                        &   $ g_i$    &   $|g_i|$          &  $g_i$           & $|g_i|$      \\
$\pi \Xi$            &  $2.11+i1.07$     &  $2.37$         & $-0.36-i0.24$    &  $0.43$  \\
$\bar{K}\Lambda$     &  $-2.10-i0.09$    &  $2.10$         & $-0.81+i0.02$  &  $0.81$   \\
$\bar{K}\Sigma$      & $-0.86+i0.38$     &  $0.94$         &   $2.26+i0.03$   &   $2.26$   \\
$\eta \Xi$           &  $-0.19+i0.12$     &  $0.23$        &  $1.04-i0.07$    & $1.04$   \\
\hline
\hline

\end{tabular}
\end{table}

\begin{figure}
\centering
\includegraphics[width=0.55\textwidth]{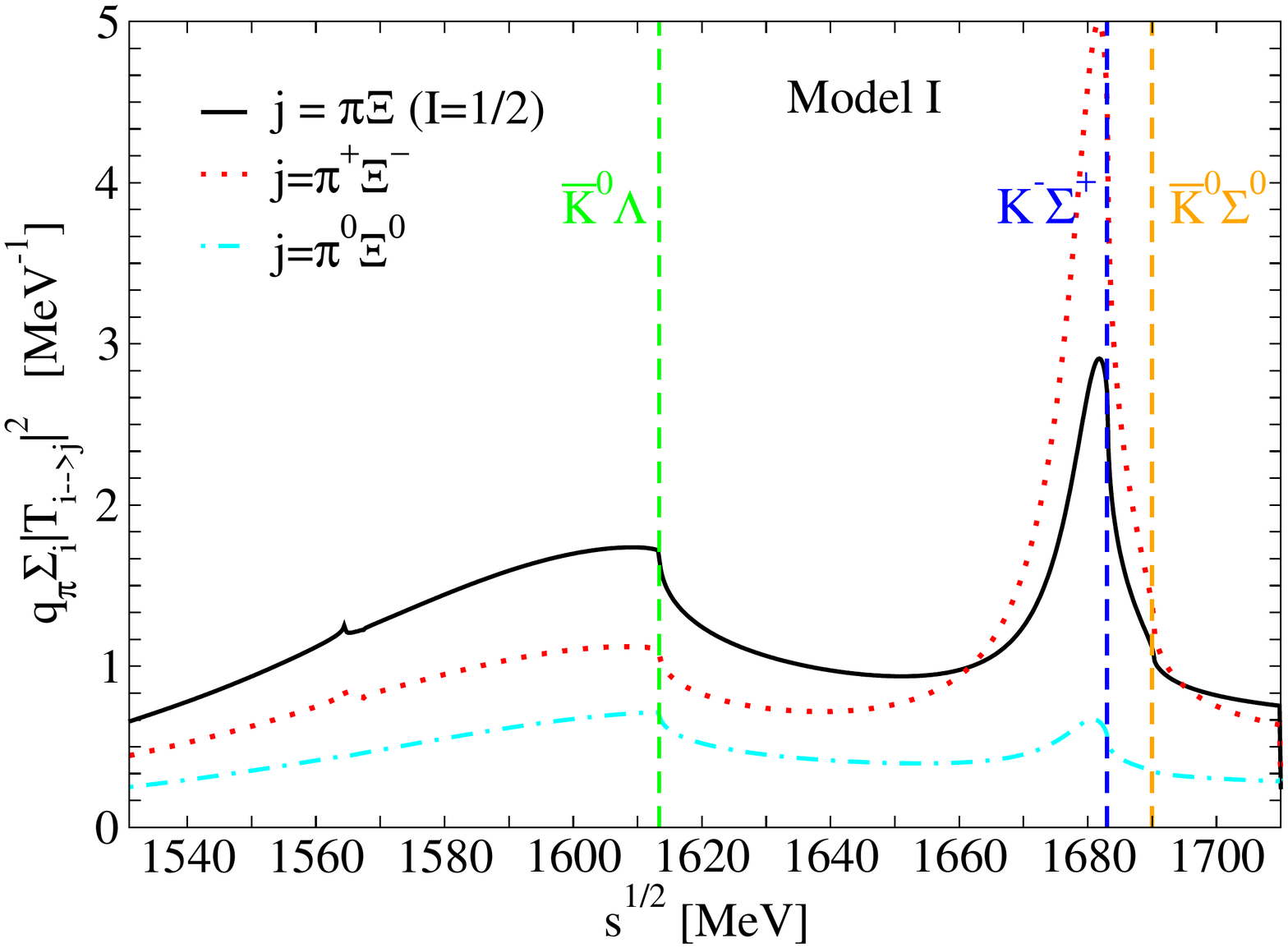}
\vspace{-0.6cm}
\caption{Sum of amplitudes squared times a phase space factor. The vertical dashed lines represent the channel threshold locations.} 
  \label{fig:t2_1}
\end{figure}

As it could be expected, because of the increased relative relevance of $\bar K$'s and $\eta$'s over $\pi$'s in the $S=-2$ sector, our system prefers a bit higher value for $f$ (see results for Model II in Table~\ref{tab:outputs_fits}). On the other hand, 
the corresponding SC barely differ from those of Model I. These combined modifications provide however a notable improvement of the pole location in the complex plane in order to describe the experimental states as seen in Table~\ref{tab:spectroscopy}. The theoretical masses reach values closer to the nominal ones, the same can be said about the theoretical $\Xi(1690)$ width. Regarding the $\Xi(1620)$ width for Model II, the new value exceeds by $20$~MeV that of Model I.

\begin{figure}
\centering
  \includegraphics[width=0.55\textwidth]{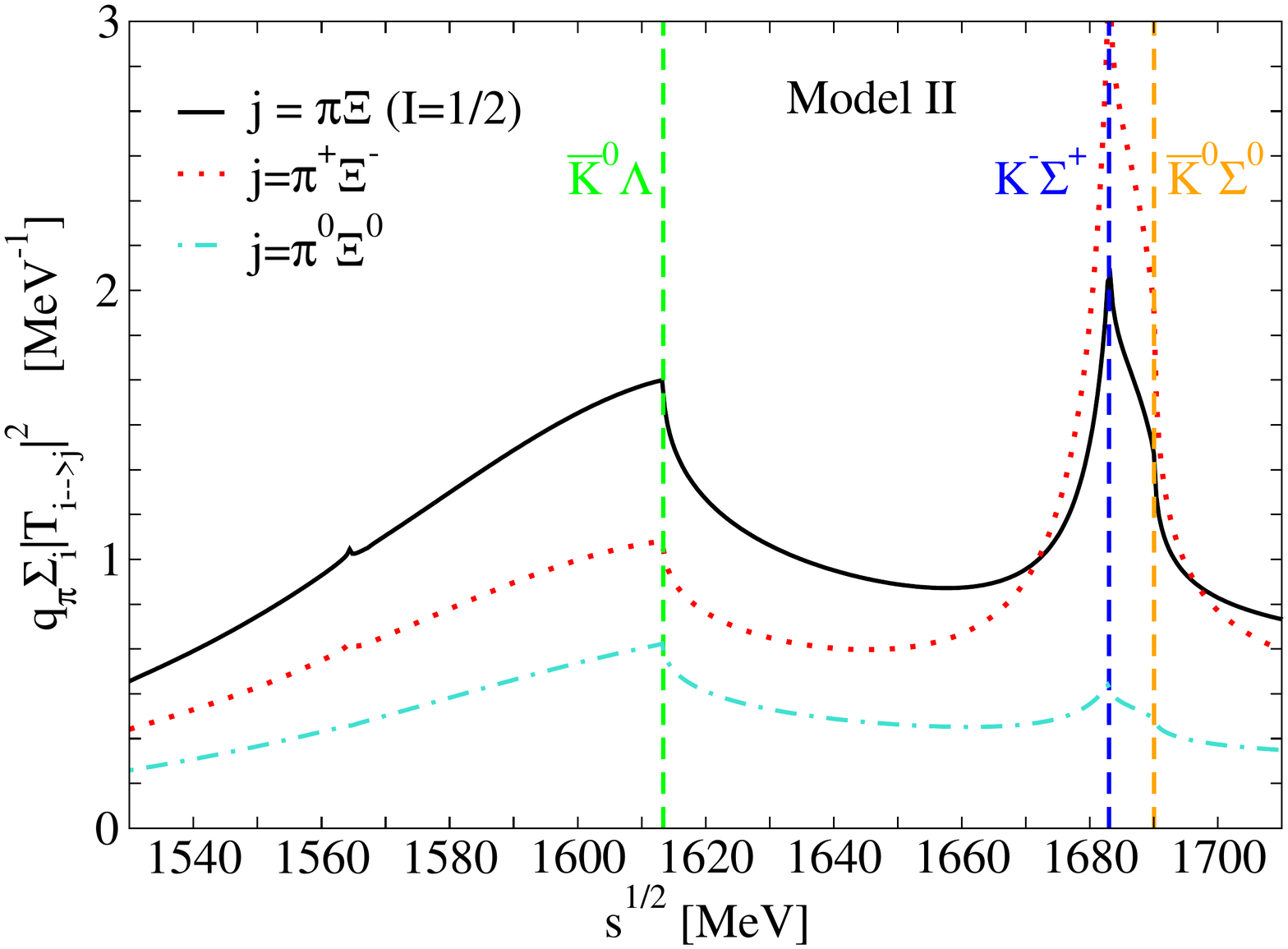}
\vspace{-0.6cm}
\caption{
Sum of amplitudes squared times a phase space factor. The vertical dashed lines represent the channel threshold locations. } 
  \label{fig:t2_2}
\end{figure}

To aid the understanding of the apparent incompatibility between the theoretical and the experimental  width of the $\Xi(1620)$ state and to give an idea of the $\pi\Xi$ spectrum that our models would provide, the quantity $q_{\pi} \mid \sum_{i} T_{i\to \pi\Xi} \mid^2$ is represented in Figs.~\ref{fig:t2_1} and \ref{fig:t2_2}, where $T_{i\to \pi\Xi}$ is the amplitude for the $i \to \pi\Xi$ transition obtained here with either Model I (Fig.~\ref{fig:t2_1}) or  Model II (Fig.~\ref{fig:t2_2}), with $i$ taking any of the six coupled channels involved in this sector and the final $\pi\Xi$ state can denote any of the physical channels: $\pi^+\Xi^-$, $\pi^0\Xi^0$. The momentum of the $\pi$ meson in the $\pi\Xi$ center-of-mass frame, $q_{\pi}$, acts as a phase-space modulator. We note that, in front of each amplitude $T_{i\to \pi\Xi}$ in the former expression, one should have included a coefficient gauging the strength with which the production mechanism excites the particular meson-baryon channel $i$. Given the qualitative character of this production mechanism, we have assumed all these coefficients to be equal. Returning to the problem of the oversized theoretical $\Xi(1620)$ width, either Fig.~\ref{fig:t2_1} or Fig.~\ref{fig:t2_2} clearly show an effective reduction of the width compared to the theoretical value because of the Flatt\'e effect \cite{Flatte:1976xu}. This is a well-known effect that takes place when a resonance is located below a threshold of a channel whose coupling to this structure is strong, and this is exactly the situation with the $\bar K^0\Lambda$ which opens $5$~MeV above the peak of the theoretical resonance. If we focus on the first structure generated by Model II, which we consider our best model, the width is around $90$~MeV that it is in much better agreement with the measured one Eq.~(\ref{eq:Xi_1620}). The second structure present in Fig.~\ref{fig:t2_2} also shows a threshold effect produced by the opening of the $K^-\Sigma^+$ that distorts the typical Breit-Wigner shape of a resonance. In any case, and despite being merely indicative and lacking of any background by construction, the spectra displayed in both Figures reproduces qualitatively well the structures present in the experimental $\pi^+\Xi^-$ spectrum shown in FIG.~2 of \cite{Belle:2018lws}.


For a general $2$-body decay, the total width of a s-wave resonance with mass $M_R$ into the ith-channel is given by 
$
 \Gamma_{R}^{i} = p_{i}\, | g_{R, i} |^2 / ( 8\pi M^2_{R})\,,
$
where the $p_i$ is the modulus of the outgoing-particle three-momentum in the parent's rest frame, and $g_{R, i}$ stands for the coupling of the resonance to the corresponding ith-channel. The $\Xi(1690)$ branching ratios, collected in Ref.~\cite{ParticleDataGroup:2022pth}, can be computed as follows:
$$
 B_1={\Gamma_{\Xi(1690)}^{\pi \Xi}}/{\Gamma_{\Xi(1690)}^{\bar{K}\Sigma}} =0.24\,, \  B_2={\Gamma_{\Xi(1690)}^{\bar{K}\Sigma}}/{\Gamma_{\Xi(1690)}^{\bar{K}\Lambda}} =1.6\,,
  \label{eq_width2}
$$
where, for $B_1$ we have employed the averaged meson and baryon mass values for the $\pi \Xi$ channels, while for $B_2$ we took $m_{\bar{K}}=m_{K^-}$ and $m_{\Sigma}=m_{\Sigma^+}$. The former assignation is due to the location of the theoretical resonance which is in between $K^-\Sigma^+$ and $\bar{K}^0 \Sigma^0$ thresholds thereby living $K^-\Sigma^+$ as the most accessible $\bar{K}\Sigma$ decay channel despite having some access to the $\bar{K}^0 \Sigma^0$ channel because the own width of the resonance. The corresponding couplings needed are taken from Table~\ref{tab:spectroscopy}, this calculation is just limited to our best model, Model II. It is interesting to notice when comparing the theoretical values of $B_1$ and $B_2$ to the experimental ones \cite{ParticleDataGroup:2022pth} that they are not only of the same order of magnitude but also, in the case of $B_2$, the branching ratio is within the range of the experimental values. Thus, we obtained a natural explanation for the $\Xi(1690)$ decay branching ratios even without including these data into a fitting procedure.

To complete presenting the results of our model, we also calculate the effective range and scattering length at $\bar{K}^0 \Lambda$ threshold and show these in Table \ref{tab4}. These values potentially can be extracted from the Correlation Function using femtoscopy techniques by ALICE collaboration, similarly to the ongoing analysis of the $K^- \Lambda$ ones \cite{ALICE:2020wvi}.

\begin{table}[h] 
  \caption{Effective range, $r_{eff}$, and scattering length, $a_0$, for $\bar{K}^0 \Lambda$ threshold.} 
  \label{tab4}
\centering
\begin{tabular}{lcc}
\hline \\[-2.5mm]
                                 & \textbf{Model I}             &  \textbf{Model II}      \\                              
\hline \\[-2.5mm]

$a^{\bar{K}^0 \Lambda}_{0}$                   & $-0.155+i\, 0.501$                         & $ -0.115 + i\, 0.495 $        \\ [1.5mm]
$r^{\bar{K}^0 \Lambda}_{eff}$           & $ -0.408 - i\, 0.413$                        & $ -0.507 - i\, 0.205  $       \\ [2.5mm]
\hline
\end{tabular}
\end{table}

\section{Conclusions}
\label{sec:conclusions}
 We have studied the meson-baryon interaction in the neutral $S=-2$ sector within an extended UChPT scheme, in which we take into account not only the leading WT term, but also the Born terms and NLO contribution for the first time in this sector. Most of the model parameters have been assumed to be $SU(3)$ symmetric and taken from the BCN model \cite{Feijoo:2018den}, where these were fitted to the large amount of experimental data in the neutral $S=-1$ sector, while the SCs have been taken as a free fitting parameters within their natural size limits.  
 
 We have shown that our model is able to generate dynamically both $\Xi(1620)$ and $\Xi(1690)$ states in a very reasonable agreement with the known experimental data. It is also important that the molecular nature of these states, provided by the present approach, naturally explains the puzzle with the decay branching ratios of $\Xi(1690)$.  Thus, once again, the reliability of the chiral models with unitarization in coupled channels and the importance of considering Born and NLO contributions for precise calculations have been proved.

\section*{Acknowledgements}
\label{sec:Acknowledgements}
The authors are very grateful to I. Vidaña and A. Ramos for their comments and careful reading of the manuscript.
This work was supported by the Ministerio de Ciencia e Innovaci\'on of Spain through the “Unit of Excellence María de Maeztu 2020-2023” award to the Institute of Cosmos Sciences (CEX2019-000918-M) and under the projects PID2020-118758GB-I00 and PID2020-112777GB-I00, and by Generalitat Valenciana under contract PROMETEO/2020/023. This project has received funding from the European Union Horizon 2020 research and innovation programme under the program H2020-INFRAIA-2018-1, grant agreement No. 824093 of the STRONG-2020 project. The work of A. F. was partially supported by the Generalitat Valenciana and European Social Fund APOSTD-2021-112.

 \bibliographystyle{elsarticle-num} 
 \bibliography{cas-refs}





\end{document}